\documentclass[aps,prl,twocolumn,showpacs,superscriptaddress]{revtex4}
\usepackage{graphicx}
\usepackage{dcolumn}
\usepackage{bm}
\usepackage{amssymb}
\usepackage{verbatim}
\usepackage{epstopdf}
\usepackage{color}

\newcommand{\beq}{\begin{equation}}
\newcommand{\eeq}{\end{equation}}

\newcommand{\eq}[1]{{(\ref{#1})}}
\newcommand{\commentout}[1]{{}}

\newcommand{\half}{\hbox{$1\over2$}}

\begin{document}

\title{Pulsating and persistent vector solitons in a Bose-Einstein condensate in a lattice upon phase separation instability}

\author{Uttam Shrestha}
\affiliation{Department of Physics, University of Connecticut,
Storrs, CT 06269-3046}
\author{Juha Javanainen}
\affiliation{Department of Physics, University of Connecticut,
Storrs, CT 06269-3046}
\author{Janne Ruostekoski}
\affiliation{School of Mathematics, University of Southampton,
Southampton, SO17 1BJ, UK}
\date{\today}

\begin{abstract}
We study numerically the outcome of the phase separation instability of a dual-species Bose-Einstein condensate in an optical lattice. When only one excitation mode is unstable a bound pair of a bright and dark soliton-like
structures periodically appears and disappears, whereas for more than one unstable mode a persistent soliton-antisoliton pair develops. The oscillating soliton represents a regime where the two-species condensate neither remains phase-separated nor is dynamically stable.
\end{abstract}

\maketitle

%{\color{red} test}

A two-component atomic Bose-Einstein condensate (BEC) is a veritable laboratory of dynamics. Experimentally observed long lifetimes of two-species condensate mixtures have prompted many studies of phase separation~\cite{Hall,Papp,Esry,LAW97,SIN99,Kasamatsu}, symmetry-breaking transitions \cite{LEE09}, spin \cite{LEW02} and shock \cite{DUT01} waves, and vector soliton structures~\cite{DUT01,RUO01,Busch,SAV03,Anderson,Becker}. Similar phenomena are also possible in nonlinear optics~\cite{Christ}. A two-species BEC may be loaded in an optical lattice~\cite{MAN03}, which should afford an even wider tunability and control of system parameters than non-lattice systems.

Phase separation is a generic phenomenon that occurs in different forms of matter and plays an important
role, e.g., in mixing of liquids.
It has previously been shown~\cite{Busch,Christ,Anderson,DUT01,Becker} that phase separation between two superfluid components in a BEC can stabilize a bright soliton that would otherwise not occur. In the phase separation regime a bright solitary wave in one component exists trapped inside a density dip of a co-propagating dark soliton in the other component. Generalizations of dark-bright solitons, such as 3D particle-like solitons in a two-species BEC \cite{RUO01,SAV03,BAT02} and analogous cosmic strings in the early Universe cosmology \cite{cos}, are
similarly stabilized by phase separation.

\begin{figure}\vspace{-0.0cm}
\includegraphics[width=0.9\columnwidth]{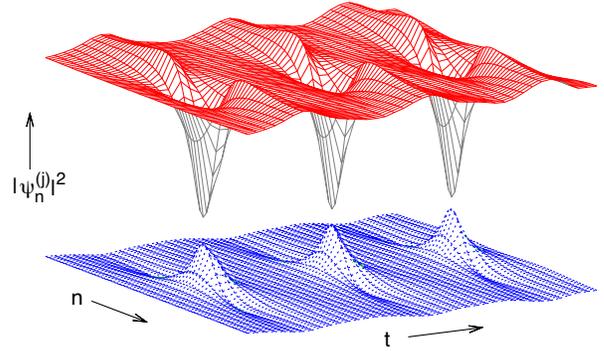}
\vspace{-0.4cm}\caption{(Color online) Creation of a pair of a bright soliton (peak) and a dark soliton (dip) resulting from a dynamical instability. The parameters are $\chi_{11}=1$, $\chi_{22}=100$, $\chi_{12}= 12$, and  $J_1=J_2=0.5$, with the lattice size $L=64$. The density of the dark soliton has been rescaled to make it visible. The arrows indicate the directions of lattice sites $(n)$, time $(t)$ and the density of each species $(|\psi_n^{(j)}|^2)$.
}
%\vspace{-1.2cm}
\label {f1}
\end{figure}

In this Letter we study theoretically the outcomes of a dynamical (modulational) phase separation instability of a two-component BEC in a lattice using the number of unstable excitation modes as the organizing principle. For one unstable mode, we surprisingly find a quasiperiodically emerging bound pair of bright and dark soliton-like structures (a soliton and an antisoliton) as the first sign of the phase separation instability (Fig.~\ref{f1}), whereas more than one unstable mode can lead to a formation of a persistent vector soliton or to more familiar violent phase mixing dynamics. The quasiperiodic soliton formation represents a regime where the system neither is stable, nor can undergo phase separation. It is reminiscent of a pulsating instability due to anomalous dispersion in a single-species BEC in a lattice~\cite{Shrestha,Shrestha1}. However, here the nature of the instability is phase separation, which cannot occur in a single BEC. We are led to speculate that an instability unfolding as pulsations is a widespread phenomenon in nonlinear systems with one unstable mode. Our proposed lattice system could be used as a testing laboratory for the pulsating instability and soliton formation; it is of particular interest that current experiments should be able to demonstrate these phenomena.

In order to study in detail the dynamics of a two-species BEC in a regime where the uniform density profile in the lattice becomes dynamically unstable we performed linear stability analysis of a classical
mean-field dynamics of the BECs along the lines of Ref.~\cite{Ruoste}. A weakly interacting BEC in a deep optical lattice can be described by the tight-binding version of the Gross-Pitaevskii equation, known as discrete nonlinear Schr\"odinger equation (DNLSE). The two-species version of DNLSE reads
\begin{equation}
i{d\over d t}\psi_{n}^{(j)}=-J_j(\psi_{n+1}^{(j)}+\psi_{n-1}^{(j)})+\sum_{k=1}^{2}\chi_{jk} |\psi_{n}^{(k)}|^2\psi_{n}^{(j)}\,,
\label{E1}
\end{equation}
where $J_{j}$ ($J_{j}>0$) denotes the hopping amplitude of the atoms between adjacent lattice sites for the species $j=1,2$, and the interaction coefficients $\chi_{jk}$ describe the onsite atom-atom interactions that
are proportional to the intra-species and the inter-species $s$-wave scattering lengths $a_{jj}$ and $a_{jk}$ ($j\neq k$), as well as to the overlap of the lowest vibrational state wave functions $\phi_{n}^{(j)}$ (the Wannier functions) of the two species in each lattice site, so that $\chi_{jj}\simeq
4\pi\hbar N_i a_{jj}/m_i \int d^3r\, |\phi_{n}^{(j)}|^4$ and $\chi_{jk}\simeq 2\pi\hbar\sqrt{N_iN_j} a_{jk}/\mu \int d^3r\,|\phi_{n}^{(j)}|^2 |\phi_{n}^{(k)}|^2$, for $j\neq k$. Here $m_j$ is the atomic  mass of the $j^{\rm th}$ BEC component and $\mu=m_1m_2/(m_1+m_2)$ is the reduced mass.
We use the normalization $\sum_{n=1}^{L} |\psi_{n}^{(j)}|^2=1$ as the number of atoms is conserved for each species separately. Throughout our calculation we consider a finite lattice length $L$, with periodic boundary conditions such that the lattice site with the index $n=0$ is equivalent to the lattice site with the index $L$. Technically we are dealing with a ring lattice, but our results clearly should have counterparts for other types of  boundary conditions.

The DNLSE~(\ref{E1}) may be solved with the standard ansatz
$\psi_n ^{(j)}=\,e^{i(p_j n- \omega_j t)}/\sqrt{L}$, providing a stationary solution with
the two chemical potentials $\omega_j=-2 J_j\cos p_j+ \sum_{k=1}^{2}\chi_{jk}/L$. Here $p_j$ is the carrier wave momentum for the $j^{\rm th}$ species. The boundary conditions quantize the $p_j$ as $p_j =P_j (2 \pi/L) $, where $P_j$ is an integer that may be chosen from the interval $[-L/2,L/2)$. For notational convenience we take the number of lattice sites $L$ to be even.

The stability of the steady state solution for the two-species BEC may be studied using the classical
Bogoliubov expansion. By substituting
\begin{equation}
 \psi_n^{(j)}= (1+u_q^{(j)}{\text e}^{iq n}-[v_q^{(j)}]^*{\text e}^{-iq n}) {\text e}^{i(p_j n-\omega_jt)}/\sqrt{L}
\label{E4}
\end{equation}
into Eq.~(\ref{E1}) and collecting the first-order terms in $u_q^{(j)}$ and $v_q^{(j)}$, which are assumedly small, we obtain a matrix equation of the type~\cite{Ruoste},
$
i{d{\xi}/ dt}={\mathfrak M}{\xi}$,
where ${\mathfrak M}$ is a $4\times 4$ non-Hermitian matrix and $\xi=[u_q^{(1)},v_q^{(1)},u_q^{(2)},v_q^{(2)}]^T$.	
The eigenvalues of ${\mathfrak M}$ correspond to the normal mode frequencies of the system. Due to the periodic boundary conditions the quasimomenta $q$ are also quantized as
\begin{equation}
q=2\pi Q/L\,,
\label{qrel}
\end{equation}
with $Q$ taking on the integer values except zero in the range  $[-\frac{L}{2},\frac{L}{2})$. A simple form for the analytic expressions for excitation frequencies $\Omega_q$ are obtained when two BECs have the same atomic currents, $J_1\sin (p_1)=J_2 \sin (p_2)$~\cite{Ruoste}, namely,
\begin{equation}
\Omega_q=t_1\pm\sqrt{\half(t_2\pm t_3)}\, ,
\label{E7}
\end{equation}
where $t_1 = 2J_1 \sin(p_1)\sin(q)$ represents a Doppler shift term of the excitation frequencies due to the superfluid flow, $t_2 = \nu_{1,q}^2+\nu_{2,q}^2$ and
\beq
t_3 = \sqrt{(\nu_{1,q}^2-\nu_{2,q}^2)^2 +16\epsilon_{1,q}\epsilon_{2,q}\chi_{12}^2/L^2 \cos p_1\cos p_2 }\, ,
\eeq
are defined in terms the single-condensate normal mode frequencies $\nu_{j,q}$ (without the Doppler
shift term)
\beq
\nu_{j,q}^2=\epsilon_{j,q}\cos(p_j) [\epsilon_{j,q}\cos(p_j)+{2\chi_{jj}/L}]\,,
\label{E8}
\eeq
and $\epsilon_{j,q}=4J_j \sin^2(q/2)$ is the spectrum of an ideal, non-moving BEC. The flow is stable if the frequencies in Eq.~(\ref{E7}) are real for all $q\ne0$, otherwise there are small excitations in the system that grow exponentially in time. For the purposes of the stability analysis, the excitation modes with the indices $Q$ and $-Q$ may, and will, be regarded as equivalent.

For a uniform single-component BEC with repulsive atom-atom interactions the condensate flow in a lattice turns dynamically unstable when the flow velocity exceeds a critical value~\cite{BUR01}. In a two-component BEC dynamical instabilities may arise both from critical velocity and from phase separation of the two species~\cite{Ruoste}. Here we study the dynamics in the normal dispersion regime, for quasimomenta $p_1,p_2<\frac{\pi}{2}$ and interaction parameters $\chi_{jk}>0$, so that the instability is ascribed exclusively to phase separation.  According to Eq.~(\ref{E7}), those lowest energy modes turn unstable for which $q$ satisfies
\beq
\chi_{12}^2>\chi_{11} \chi_{22}+q^2 L\left(j_2\chi_{11}+j_1\chi_{22}\right)/2+q^4 L^2 j_1 j_2/4 \, .
\label{E11}
\eeq
Here we are studying the lowest excitation modes with $0<Q\ll L$, where $q$ is given by Eq.~(\ref{qrel}), and write $j_k\equiv J_k \cos p_k$. The number of unstable modes may be obtained from Eqs.~\eq{qrel} and~\eq{E11} by solving for $Q$. In the limit of a large number of lattice sites $L\rightarrow \infty$, we obtain the condition for instability $\chi_{12}^2>\chi_{11} \chi_{22}$ \cite{Ruoste}. This is an obvious counterpart of the iconic condition for the instability of free-space condensates with equal-mass atoms~\cite{Esry,LAW97}, $a_{12}>a_{11}a_{22}$. A similar analysis may also be performed in the case that one of the BECs exhibits anomalous dispersion due to superfluid flow, but the resulting stability conditions are significantly altered \cite{Ruoste}.

We next simulate the dual-species dynamics in a lattice using a sixth-order accurate split-step FFT method as described in~\cite{Javanainen} for a single-condensate Gross-Pitaevskii equation. We start with a homogeneous non-moving mixture of two BECs, adding a small amount of random noise ($\sim10^{-6}$) to the condensate amplitudes to speed up the instability. We first concentrate on what we call weak instability, when only one excitation mode with $Q=1$ is unstable, and also assume that the interaction strengths satisfy $\chi_{22}\gg\chi_{11}$.  Because of the proliferation of parameters, we do not wish to use any of them in particular as a scale of dimensional quantities. Instead, we have arbitrary compatible units for energy, frequency, and time, so that dimensional quantities appear as dimensionless numbers in our further analysis.

In Fig.~\ref{f1} we show the time evolution of the density profiles $|\psi_n^{(j)}|^2$ of initially non-moving BECs as the instability runs its course. The parameter values are $\chi_1=1,~\chi_2=100$ and $\chi_{12}=12$, for $L=64$ and $J_1=J_2=0.5$. The critical value of inter-species interactions for the onset of the instability for these parameters is $\chi_{12}=10.75$, according to Eq.~(\ref{E11}). The value $\chi_{12}=12$ in Fig.~\ref{f1} corresponds to one unstable mode, as the threshold for the second mode to become unstable is $\chi_{12}=12.75$. Here one expects phase separation, and a region with a pronounced perturbation between  the densities (numbers of atoms per lattice site) of the two species indeed  develops. However, Fig.~\ref{f1} shows that upon further evolution the species mix again. The system returns close to its initial state, subsequently an abortive phase separation occurs again, and so on.

In this feature the system bears a strong resemblance to the previously demonstrated pulsating instability  that is associated with superfluid flow~ \cite{Shrestha,Shrestha1}. The mechanism is also likely to be as we surmised before: At the point of instability the stationary state, which used to be an elliptic fixed point of the dynamics, turns into a hyperbolic fixed point. In the process a homoclinic orbit emerges that starts from the hyperbolic fixed point in the unstable direction and returns to the same fixed point in the direction of a corresponding excitation mode that shrinks exponentially. Starting in the vicinity of the hyperbolic fixed point, the system approximately tracks the homoclinic orbit, which leads to periodic recurrences of phase separation and the initial state.

It may seem counterintuitive that the minority species, the one with much weaker interspecies interactions, may form a local density maximum in spite of the repulsive interactions and normal dispersion, but there is a well-known analog in the literature in the form of a bound pair of bright and dark solitons both in free space~\cite{Christ} and in a harmonic trap~\cite{Busch,Anderson}.  In what is ultimately a case of phase separation, the background of the coincident dark soliton stabilizes the bright soliton, providing an effective trap due to the repulsion between the two species when the variation in the total atom density is energetically less favorable than in relative atom densities \cite{SAV03,BAT02}. The pulsating variations of atom density we have described thus far neither constitute an energy minimum nor are they stationary, but on the strength of the analogy we nevertheless refer to our finding as a bound pair of a dark and a bright soliton.

\begin{figure}
\includegraphics[width=0.9\columnwidth]{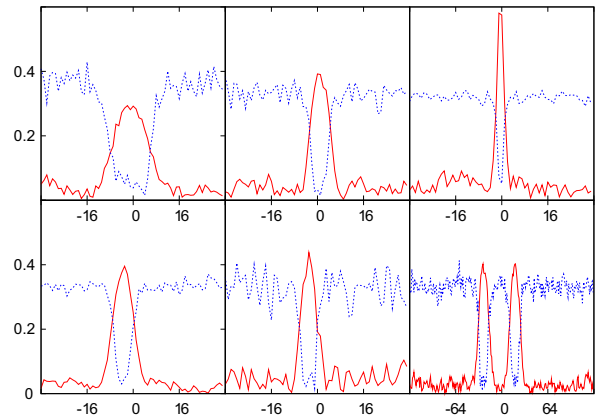}
\vspace{-0.2cm}
\caption{
(Color online) Top row from left: Profiles of persistent soliton pairs for $(\chi_{11},\chi_{22},\chi_{12})=(1,10,8)$, $(1,100,20)$, and $(1,1000,60)$ for the common parameters $J_1=J_2=0.5$, $L =64$. Here each example starts with four unstable modes. Bottom row from left: Persistent soliton pairs for $\chi_{12}=15$ (two unstable modes), $36$ (eight unstable modes), and $20$ (eight unstable modes). Here we have $J_1=J_2=0.5$, $\chi_{11}=1$, $\chi_{22}=100$, and $L =64$, except in the last case $L =256$. The dip in the antisoliton (dashed line) has been multiplied by a factor of 2.5 to enhance
the visibility. The $x$ axis in each panel counts the lattice sites whereas the $y$ axis is for the densities of minority (full line) and majority (dashed line) species.
\vspace{-0.7cm}}
\label {f3}
\end{figure}

The large ratio $\chi_{22}/\chi_{11}$, e.g., due to population imbalance, is essential for the emergence of solitons. For nearly equal nonlinearities we do not observe recognizable localized structure, and the structure narrows with the increasing ratio $\chi_{22}/\chi_{11}$.  On the other hand, having more than one excitation mode dynamically unstable eventually leads to persistent phase separation, for $\chi_{22}/\chi_{11}\gg1$ in the form of a dark-bright soliton, with the associated phase kink in the dark soliton. We illustrate this in Fig.~\ref{f3} by showing the density profiles of the final soliton pair for three different ratios of $\chi_{22}/\chi_{11}$ (top row). In each case the number of initially unstable modes was four. The persistent soliton profiles also become narrower when the number of unstable modes increases; Fig.~\ref{f3} (bottom row). A large number of unstable modes can result in the two species being mixed in a more complicated way with no recognizable solitons \cite{latticemixing}, analogously to the previous experiments \cite{Papp}. Moreover, we found that, in the case of several unstable modes, increasing the number of sites leads to more solitons, as also demonstrated in Fig.~\ref{f3}. For balanced populations we may also observe maximally phase-separated density profiles where the two species occupy the separate halves of the lattice \cite{Rapti}. All soliton profiles in Fig.~\ref{f3} are surrounded by sound wave excitations appearing as a fluctuating background density. While these soliton pairs appear stable, there is nothing in our translationally invariant model to pin them, so they jitter around.

\begin{figure}
\includegraphics[width=0.9\columnwidth]{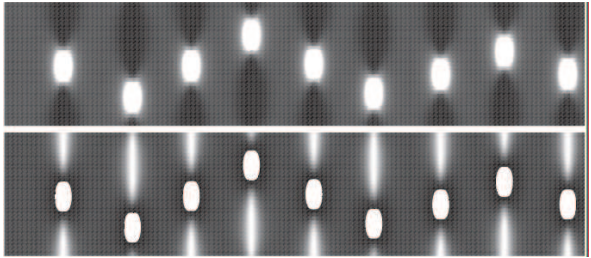}
\vspace{0.0cm}\caption{ Motion of bright (top) and dark (bottom) pulsating solitons in an added harmonic potential $V_n=\frac{1}{2}\kappa (n-\frac{L}{2})^2$. Lattice position {from a total of 64 sites}  (vertical) is shown as a function of time (horizontal).  Except for the harmonic trap with $\kappa=10^{-6}$, the parameters are the same as in Fig.~\ref{f1}.  Brighter shading represents higher densities in the top panel, and lower densities at the bottom.}
\label {f2}
\vspace{-0.5cm}
\end{figure}

This dark-bright soliton composite can be viewed as a bound state of two particles, a {vector soliton}, as is seen by setting it in motion. Here we resort to  numerical  experiments inspired by oscillatory soliton dynamics that has been both discussed theoretically~\cite{Busch} and observed experimentally~\cite{Becker}. Initially we have a homogeneous plane wave with $p_j=0$ for each component. We switch on a weak trapping potential in the axial direction, $V_n=\frac{1}{2}\kappa (n-\frac{L}{2})^2$ with $\kappa=10^{-6}$, and wait for the pulsating soliton pair to emerge. For the parameter values we have used it always comes at the trap center, and stays there if the trap remains on unchanged. However, once we have the soliton pair, we abruptly displace the center of the trap, whereupon the trapping forces make the solitons oscillate back and forth. For the pulsating soliton pair, the oscillation amplitude in the trap can be much larger than the initial displacement. This is demonstrated in Fig.~\ref{f2} where the trap center was displaced by one lattice site, yet we see oscillation amplitudes of about 20 sites. The bright and dark solitons in a pulsating pair stay together while they oscillate back and forth just like the components of a persistent vector soliton, but for a persistent soliton the oscillation amplitudes are comparable to the initial displacement.

In the case of a single species, weak instability requires a weak nonlinearity. The pulsating instability in a single-component superfluid flow~\cite{Shrestha} therefore entails long time scales, and the attendant experimental problems. In contrast, the present pulsating soliton-antisoliton pair seems to be within easy reach of current experimental techniques. This is because the two species stabilize each other, and the interactions can be large while the system remains in the regime of weak instability with just one unstable excitation mode. The time scale of the instability may thereby be shortened and, e.g., is $\tau=1/|\Im \Omega_q| \simeq 80$ for the parameters of Fig.~\ref{f1}. If the root energy scale is the recoil energy, as is reasonable for optical lattices, for ${}^{87}$Rb this translates to 20$\,$ms. Continuing with our qualitative example, the time between two consecutive pulses in Fig.~\ref{f1} would be 0.4$\,$s.

One could also think of in-principle problems such as failure of the DNLSE to approximate well the underlying quantum field theory. However, the DNLSE only fixes the nonlinearities $\chi_{kl}\propto \sqrt{N_k} \sqrt{N_l}$ and, for a given values of $\chi_{kl}$ the classical mean-field theory in an optical lattice becomes increasingly accurate for higher atom numbers. In our example estimates of quantum effects on the pulsating instability in a single-species BEC we have found that the DNLSE may fare well for atom numbers as small as hundreds~\cite{Shrestha1}.

In conclusion, we have described theoretically a feature that uniquely combines phase separation, soliton-antisoliton pair, and pulsating instability: In the limit of only one unstable excitation mode a bound pair of bright and dark solitons that emerges and disappears nearly periodically while a dynamical instability of a two-species BEC in an optical lattice runs its course. Basically, we have repeated attempts at phase separation that do not quite make it. However, if more than one excitation mode is unstable, phase separation goes through and produces a persistent soliton-antisoliton pair. The dual-species condensate in an optical lattice  also exhibits other instabilities besides phase separation, and additional peculiar behaviors are likely when these instabilities take off.

We acknowledge financial support from NSF (PHY-0750668) and EPSRC.

\end{document}